# AI-Powered Hybrid Intrusion Detection Framework for Cloud Security Using Novel Metaheuristic Optimization

Maryam Mahdi Alhusseini [*1],
(Member, IEEE)
[1] Middle Technical University, Polytechnic College of Engineering - Baghdad, Baghdad, Iraq,
mariammahdi@mtu.edu.iq
, maryammalhusseini@ieee.org

Alireza Rouhi [2]
[2] Faculty of Information Technology and Computer Engineering, Azarbaijan Shahid Madani University, Tabriz, Iran
rouhi@azaruniv.ac.ir

Mohammad-Reza Feizi-Derakhshi [3]
[3] Computerized Intelligence Systems Laboratory, Department of Computer Engineering, University of Tabriz, Tabriz, Iran
mfeizi@tabrizu.ac.ir

*Corresponding Author: Maryam Mahdi Alhusseini*

**ABSTRACT**
Cybersecurity poses considerable problems to Cloud Computing (CC), especially regarding Intrusion Detection Systems (IDSs), facing difficulties with skewed datasets and suboptimal classification model performance. This study presents the Hybrid Intrusion Detection System (HyIDS), an innovative IDS that employs the Energy Valley Optimizer (EVO) for Feature Selection (FS). Additionally, it introduces a novel technique for enhancing the cybersecurity of cloud computing through the integration of machine learning methodologies with the EVO Algorithm. The Energy Valley Optimizer (EVO) effectively diminished features in the CIC-DDoS2019 dataset from 88 to 38 and in the CSE-CIC-IDS2018 data from 80 to 43, significantly enhancing computing efficiency. HyIDS incorporates four Machine Learning (ML) models: Support Vector Machine (SVM), Random Forest (RF), Decision Tree (D_Tree), and K-Nearest Neighbors (KNN). The proposed HyIDS was assessed utilizing two real-world intrusion datasets, CIC-DDoS2019 and CSE-CIC-IDS2018, both distinguished by considerable class imbalances. The CIC-DDoS2019 dataset has a significant imbalance between DDoS assault samples and legal traffic, while the CSE-CIC-IDS2018 dataset primarily comprises benign traffic with insufficient representation of attack types, complicating the detection of minority attacks. A downsampling technique was employed to balance the datasets, hence improving detection efficacy for both benign and malicious traffic. Twenty-four trials were done, revealing substantial enhancements in categorization accuracy, precision, and recall. Our suggested D_TreeEVO model attained an accuracy rate of 99.13% and an F1 score of 98.94% on the CIC-DDoS2019 dataset, and an accuracy rate of 99.78% and an F1 score of 99.70% on the CSE-CIC-IDS2018 data. These data demonstrate that EVO significantly improves cybersecurity in Cloud Computing (CC).

## 1 Introduction

Cloud computing (CC) has revolutionized how organizations handle data, offering unparalleled adaptability and effectiveness [1]**.** This is mainly accomplished via virtualization, which permits the operation of several Virtual Machines (VMs) on one physical server and facilitates seamless VM migration between different hosts. These capabilities offer significant advantages, including enhanced hardware use, resource protection, and isolation. Nonetheless, dependence on distributed systems and environments with multiple tenants brings substantial security challenges. Virtualization, along with different cloud technologies, presents distinct security vulnerabilities that are drawing increasing interest from cybercriminals [1]. Virtualization, in conjunction with other cloud-related technologies, poses security problems. Cloud computing has revolutionized organizational data management, but with its prevalence, robust cybersecurity measures are crucial. Cyber threats pose significant risks to CC. systems' security [2].

Cybersecurity presents major hurdles for security professionals because of the rising number, variety, and regularity of cyberattacks and detrimental activities [3, 4]. Handling large datasets that contain many attributes is a considerable challenge in machine learning, as high-dimensional data can negatively affect model efficacy, increase computational costs, and make feature selection more difficult. Machine learning methods find it challenging to handle the extensive number of input features, which poses a significant issue for researchers [3]. Data preparation is necessary for machine learning algorithms. Machine learning requires feature selection, a common data preparation approach [4].

Feature selection (FS) is the process of identifying relevant characteristics or a potential subset of features. The evaluation criteria are employed to obtain an optimal subset of features [5, 6, 7]. In settings with high-dimensional data, where there are considerably fewer samples than features, selecting the right subset of features can be quite difficult [6]. The imperative for proficient management of high-dimensional data has resulted in the creation of feature selection techniques to remove irrelevant and redundant attributes, hence enhancing generalization and reducing noise. The Moth Flame Optimization (MFO) algorithm has been utilized for Feature Selection (FS) [8, 6], resulting in optimized input data and improved classifier accuracy. Similarly, heuristic methodologies using algorithms like the Firefly Algorithm (FA) and the Ant Lion Optimization Algorithm (ALO) have shown promise in the advancement of efficient IDSs [9].

Metaheuristic algorithms are essential for feature selection in high-dimensional datasets because they can explore complex solution spaces and avoid local optima. Energy Valley Optimizer (EVO) [10, 6], Ant Colony Optimization, Particle Swarm, and Genetic Algorithms, are popular feature selection techniques because of their efficiency and versatility. Intrusion attacks rank among the most prevalent kinds of cyber assaults [4]. Consequently, there is a significant demand for an intrusion detection system to effectively identify and alleviate their effects. Security challenges are a fundamental requirement of modern computer systems, having recently acquired considerable importance due to the substantial increase in both the frequency and intensity of malicious attacks. Anomaly detection is a fundamental element of intrusion detection systems that enables the recognition of new forms of security attacks [11].

Intrusion Detection Systems (IDSs) are crucial in mitigating threats linked to cloud computing. In the realm of cloud computing, which predominantly depends on the Internet, a range of security solutions, particularly IDSs, is essential for identifying attackers and malware [6]. A system for detecting intrusions is a software or hardware tool that observes both internal and external network activities to spot unusual behavior that may signal a possible cyber-attack, alerting the administrator of the network or system. The ability of an IDS to detect both known and novel threats makes it an outstanding choice for safeguarding cloud computing environments [12, 13]. Intrusion Detection Systems (IDSs) are the key to eliminating system attacks [14, 15]. Malicious intrusions are a significant factor in various network security incidents, posing substantial threats to both network users and organizations [16]. IDSs are vital for safeguarding the security of hosts and systems[15]. Nonetheless, they encounter obstacles like biased datasets, subpar performance of classification models, and challenges in identifying anomalies. An imbalanced distribution of records within datasets results in distorted classification outcomes [4].

Downsampling and up-sampling address data imbalance [4, 15]. Downsampling improves the accuracy of the model in distinguishing between benign and harmful signals. The detection rate for a class with a limited number of instances is significantly lower than that of a class with a larger number of occurrences [6]. Arbitrarily eliminating data segments attains equilibrium in stochastic downsampling. Nonetheless, category-balanced data may yield inadequate classification outcomes as it overlooks partially overlapping categories. Enhancing the datasets employed for training and evaluating these security systems requires a focus on upgrading defensive mechanisms[16]. Augmented datasets significantly enhance the effectiveness of both offline and online intrusion detection methods. The presented methods aim to resolve the current challenges in anomaly identification stemming from insufficient natural patterns in the training data [17, 18]. Some studies utilize feature mapping to improve category discrimination, albeit this may compromise the integrity of the original dataset [19].

To address these challenges, this paper proposes a Hybrid Intrusion Detection System (HyIDS), a novel system leveraging machine learning and optimization techniques [20, 21, 22, 6, 23]. HyIDS adopts a two-stage approach to enhance intrusion detection performance in cloud computing [6].

Improving the datasets utilized for training and testing these security solutions necessitates an emphasis on advancing defense systems. Enhanced datasets substantially increase the efficacy of both offline and online intrusion detection techniques. Solutions are proposed to address the existing issues in anomaly identification due to inadequate natural patterns in the training data.

In the initial phase of this methodology, the Energy Valley Optimizer (EVO) formulates an ideal feature set using Feature Selection (FS). This feature selection approach greatly improves the system's capacity to effectively represent data and enhance intrusion detection precision.

In the second stage, HyIDS integrates four prominent machine learning models (Support Vector Machine, Random Forest, D Tree, and K-Nearest Neighbors) to achieve comprehensive and accurate intrusion detection. The combination of these models significantly improves HyIDS's ability to detect intrusions effectively. Extensive evaluations using actual datasets (CIC-DDoS2019 and CSE-CIC-IDS2018) have been conducted across various scenarios, leading to a total of 24 separate experiments. The system's performance evaluation takes into account detection accuracy, false positive rates (FPR), and computational efficiency. Early results underscore the remarkable effectiveness of the EVO algorithm in selecting features, producing a suitable feature set for HyIDS. This process of feature selection plays a crucial role in improving the overall effectiveness of the system in detecting intrusions. *The main contributions of this paper are as follows:*

1. The creation of a new hybrid method that employs the Energy Valley Optimizer (EVO) for feature selection, greatly improving the system's capacity to accurately depict the data.
2. The launch of HyIDS, which integrates EVO for feature selection with machine learning models to boost cybersecurity performance in cloud computing, particularly focusing on how feature selection influences overall system efficacy.
3. A novel application of the Energy Valley Optimizer (EVO) for feature selection results in substantial feature reduction and improved performance, choosing 43 out of 80 features from the CSE-CIC-IDS2018 and 38 out of 88 features from the CIC-DDoS2019 dataset.
4. The implementation of a downsampling technique to balance the CIC-DDoS2019 and CSE-CIC-IDS2018 datasets.
5. An investigation into the capabilities of SVM, RF, D_Tree, and KNN models within Intrusion Detection Systems
6. The enhancement of machine learning model performance using the Energy Valley Optimizer (EVO) to increase detection accuracy and bolster system resilience.
7. The effectiveness of our proposed system was assessed against existing studies on intrusion detection, demonstrating significant improvements in its performance.

The organization of the paper is as follows: Section 2 reviews relevant literature, Section 3 provides a comprehensive overview of HyIDS, and Section 4 wraps up the paper and suggests avenues for future research.

## 2 Related work

This segment emphasizes the latest progress in machine learning and optimization techniques applied to Intrusion Detection Systems (IDSs) within the realm of cloud cybersecurity.

Zhao et al. [10] introduced a hybrid IDS architecture called CFS-DE, which targeted effective feature selection to lessen dimensionality. To enhance classification performance, they employed a Weighted Stacking method that amplified the impact of robust classifiers while diminishing the influence of less effective ones. Their experiments utilizing the NSL-KDD and CSE-CIC-IDS2018 datasets yielded remarkable results, achieving accuracies of 99.87% and 87.44%, respectively. Their approach surpassed that of conventional machine learning models. In a related research effort, Joloudari et al. [24] examined five classifiers—SVM, Bayesian, Random Forest, Neural Network, and a PSO-optimized SVM. They applied the ELTA process to identify crucial predictive features, especially in the domain of liver disease detection, with the goal of enhancing classification accuracy.

Meta-heuristics, highlighted by Mirjalili et al. [25], offer effective optimization methods for real-world problems, avoiding local optima and thoroughly exploring complex search spaces. Bakro et al. [26] indicated that the protection of wireless sensor networks (WSNs), especially in IoT applications, is dependent on efficient monitoring systems, including Bluetooth-enabled sensor networks combined with motion sensors and microcontrollers. Likewise, Intrusion Detection Systems (IDS) rely on sophisticated monitoring and feature selection methods to improve detection precision. Sangaiah et al. [27] A Hybrid Ant-Bee Colony Optimization approach was suggested to enhance feature selection issues by framing them as optimization challenges. Although cloud computing environments provide numerous Internet services, security remains a significant issue. In these contexts, the defense mechanism HyIDS was employed. Several parameters assess IDSs, with the feature selection (FS) method utilized for classifying harmful and legitimate activities being the most critical factors.

The research on intrusion detection systems in cloud computing highlights various technical shortcomings. Enhancements in dimensionality reduction methods are necessary to achieve a better compromise between computational efficiency and detection precision. Additionally, numerous models exhibit decreased effectiveness when used in practical situations, suggesting a requirement for hybrid IDS methods that provide reliable accuracy across different datasets while improving feature selection and computational efficiency. Recent literature trends show an increasing adoption of feature selection techniques in IDS approaches, as summarized in Table 1.

**Table 1.** Comparative Analysis of Models and Challenges in IDS Using Benchmark Datasets

| Ref. | Datasets | Challenges | Suggestion | Model | Accuracy |
|---|---|---|---|---|---|
| R. Gautam et al. [28] | KDDCup-99 | • NIDS, HIDS<br>• Imbalanced datasets | Normalization.<br>Feature selection<br>Ensemble approach | Naïve Bayes, Adaptive boost, PART (Partial DT) | • 92.78%<br>• 97.85%<br>• 99.96% |

| Ref. | Datasets | Challenges | Suggestion | ML Models/algorithm | Accuracy |
|---|---|---|---|---|---|
| H. Xu et al. [29] | CIC-DDoS2019 | • Analyzing intricate, high-dimensional, and extensive traffic datasets. <br>• low detection accuracy | SCADE model, CNN, BiLSTM, CNN-BiLSTM-Attention model | • KNN <br>• RF | • 80.73% <br>• 88.36% |
| W. Wang et al. [30] | • KDD Cup 99. <br>• NSL-KDD | Deep learning is used to automatically extract critical features and improve detection performance | • data transformation and standardization, feature dimensionality reduction, Reconstruction Robustness | A contractive auto-encoder organized in a stacked manner (SCAE) along with an SVM model. | 75% |
| H. Jazi et al. [31] | ISCX dataset | •DDoS attack <br>•DoS attacks | Sampling techniques | Web server, nonparametric CUSUM algorithm. | 92% |
| M. Khan et al. [32] | CSE-CIC-IDS2018 | Heterogeneity of big data. | spark MLlib, conventional machine classifiers, Conv-AE | • SVM <br>• DT <br>• RF <br>• Conv-AE | 89% |

## 3 Hybrid Intrusion Detection System (HyIDS)

The research surrounding Intrusion Detection Systems (IDS) in Cloud environments identifies various technical shortcomings. There is a need for improved optimization in dimensionality reduction methods to achieve a better compromise between computational efficiency and detection accuracy. Additionally, numerous models exhibit diminished performance in practical applications, highlighting the necessity for hybrid IDS strategies that can maintain consistent accuracy across different datasets while also enhancing feature selection and computational efficiency. Tables 2 and 3 provide an overview of literature specifically related to HyIDS, indicating a growing interest in leveraging optimization algorithms and machine learning models to enhance cloud computing security.

**Table 2.** Summary of the literature on HyIDS

| Ref. | Datasets | Challenges | Suggestion | ML Models/algorithm | Accuracy |
|---|---|---|---|---|---|
| I. Aljamal et al. [22] | UNSW-NB15 | • Signature-based detection <br>• Anomaly-based detection | • Dimensionality reduction <br>• Data normalization <br>• Clustering and labelling the dataset. <br>• PCA. <br>• HyIDS | The main model is K-means, and SVM models had a different number of clusters. K-means, SVM = 10. K-means, SVM = 16. K-means, SVM = 32. K-means, SVM = 45. K-means, SVM = 64. | K-means -10 = 86% <br>K-means -16 = 85.5% <br>K-means -32 = 80% <br>K-means -45 = 87% <br>K-means -64 = 88.6% |
| R. Zhao et al. [10] | • CSE-CIC-IDS2018 <br>• NSL-KDD | • Restrict the size of the features' dimensions <br>• . HyIDS utilizing a CFS-DE algorithm combined with a weighted Stacking classification. | • CFS-DE feature selection <br>• Weighted Stacking classifier. <br>• 5-fold cross-validation <br>• HyIDS | RF, KNN, XGBoost Logistic Regression (LR) | NSL-KDD <br>RF=86.51% <br>KNN=85.70% <br>XGBoost=86.53% <br>CSECICIDS2018 <br>RF=98.00% <br>KNN=98.89% <br>XGBoost=99.05% |
| J. Zhang et al. [33] | KDD'99 | • Misuse detection <br>• Anomaly detection <br>• NIDS | • Feature selection <br>• Down-sampling <br>• HyIDS | Random Forest | •High DR: The overall DR of the hybrid system is 94.7% <br>•Low FPR: The overall FPR is 2% |
| G. Gebremariam et al. [34] | •UNSW_NB15 <br>•CICIDS2017 <br>•NSL-KDD | advanced IDS based on hybrid ML (AIDS-HML) | • Hybrid random forest <br>• Extreme gradient boost | • CLK-M <br>• Extreme gradient boost (RF-XGB) | • 100% <br>• 99.80% |
| Bakro et al. | CIC_DDoS2019 | Cloud HyIDS | GOA-GA | Random Forest | 99.97% |

| [26] | | | | (RF) | |
|---|---|---|---|---|---|
| Zhou et al. [19] | • NSL-KDD, • KDDCUP9 • UNSWNB15. | • Imbalanced datasets • HyIDS framework based on DNN | • Harris Hawk Optimization (HHO) for FS • KNN-DDAE | • DNN • HHO-DNN | • For NSL-KDD 86.79% • For KDDCUP9 94.03% • For UNSWNB15 81.92% |

**Table 3.** Summary of various (Hybrid Approaches) from prior studies

| Ref. | Security strategy | Research Outcomes |
|---|---|---|
| **Y. K. Saheed et al. [5]** | A novel hybrid feature selection (FS) approach that combines the Bat metaheuristic algorithm with the Residue Number System (RNS) has been developed. | Combining RNS with the Bat algorithm boosts detection, accuracy, F-score, and doubles IDS processing speed. |
| **R. Regan et al. [35]** | The enhanced hybrid security model | A hybrid optimized technique is proposed for detecting malicious assaults utilizing a hybrid secured model. |
| **A. Davahl et al. [13]** | Hybrid optimization and secure clustering protocol | Improves efficiency by employing a hybrid secure clustering protocol along with k-means clustering to identify attacks. |
| **S. Saif et al. [36]** | A HyIDS method that uses feature selection (FS). | Decision trees and Genetic Algorithms efficiently identified healthcare IoT vulnerabilities, reducing costs. |
| **A. Vinitha et al. [37]** | Cat Salp Swarm Algorithm based on Taylor | LEACH enables energy-efficient multi-hop routing with accurate measurement of throughput, energy, and delay. |
| **S. K. Gupta et al. [38]** | A hybrid neural network based on deep learning (DL). | Efficient intrusion detection using a hybrid Chicken Swarm–Genetic Algorithm in IoT networks. |

## 4   Energy Valley Optimizer

The Energy Valley Optimizer (EVO), proposed by Mahdi Azizi et al. in January 2023, is an innovative metaheuristic algorithm derived from sophisticated physics concepts [39]. The EVO is a modern metaheuristic inspired by particle physics, focusing on neutron–proton ratios and stability. It balances exploration and exploitation through dynamic search loops, offering precision and adaptability [39]. Despite its computational complexity, EVO benefits from modern hardware and software to optimize performance [39]. EVO incorporates a dynamic configuration for exploration and exploitation, employing three novel position vectors throughout its search process [39, 6]. The core principle of EVO is that each particle in the search space represents a potential solution, with the algorithm aiming to discover the most stable (optimal) solution through the iterative evolution of the particles. The intricacy, along with its mathematical underpinnings, establishes EVO as a potential instrument for optimization tasks, especially in high-dimensional datasets [39]. Algorithm 1 highlights our proposed pseudocode for the Energy Valley Optimizer utilized in Feature Selection.

### 4.1   Mathematical Formulation of Energy Valley Optimizer (EVO)

The Energy Valley Optimizer (EVO) is a nature-inspired metaheuristic that replicates the behavior of unstable particles seeking stability via different decay processes (alpha, beta, gamma). In this search space, each particle symbolizes a possible solution and progressively adjusts its position towards achieving optimal stability by employing equations inspired by decay [39].

1. **Alpha decay**
   Modifies the particle in accordance with the optimal solution $X_{BS}$.
   $$X_i^{NEW1} = X_i (X_{BS} (X_t^{i'})) \quad (1)$$

2. **Gamma decay**
   Revisions utilizing a nearby solution $X_{NG}$
   $$X_i^{NEW2} = X_i (X_{NG} (X_t^{i'})) \quad (2)$$

3. **Beta decay1**
   Includes the optimal particle along with the population center XCP, adjusted according to the stability level of the particle $S_{Li}$.
   $$X_i^{NEW1} = X_i + (\tau_1 * X_{BS} - \tau_2 * X_{CP})/SL_i \quad (3)$$

4. **Beta decay2**
   Promotes interaction and exchange among the top and adjacent particles.
   $$X_i^{NEW2} = X_i + (\tau_3 * X_{BS} - \tau_4 * X_{NG}) \quad (4)$$

EVO strikes a harmony between exploration and exploitation by simulating the processes of physical decay. It can be improved by combining it with other metaheuristic approaches, like SNAKE, to optimize hyperparameters such as the learning rate and the number of neurons in deep learning-based intrusion detection systems [6, 10].

### 4.2 Cost Function in Energy Valley Optimizer (EVO)

The cost function used in EVO depends on the specific optimization problem at hand. When it comes to Intrusion Detection Systems (IDS), the cost function is typically formulated based on classification accuracy, F1-score, or a combined weighting of various performance metrics.

$$\text{Cost function} = w_1 \cdot (1-\text{Accuracy}) + w_2 \cdot \text{False Positive rate} + w_3 \cdot \text{False Negative Rate} \quad (5)$$

Where:
W1, W2, and W3 are weighting factors that prioritize different aspects of classification performance.

## 5 Methodology

This section delineates our approach for the advanced Hybrid Intrusion Detection System (HyIDS), tailored specifically to fortify the cybersecurity of cloud computing environments. The architecture of the proposed HyIDS is elucidated in Section 5.1, accentuating its pivotal components and operational mechanisms. Subsequently, in Section 5.2, We present our findings and engage in discussions, ultimately assessing the outcomes of the proposed HyIDS by utilizing two well-known datasets: CIC-DDoS2019 and CSE-CIC-IDS2018. This section provides insights into the effectiveness and performance of the HyIDS in detecting and mitigating intrusions within cloud computing environments. The proposed HyIDS is a hybrid system that combines Machine Learning (ML) techniques with metaheuristic optimization algorithms. Additionally, the system incorporates feature selection (FS) to enhance classification accuracy, as supported by previous studies [11 ،27 ،4 ،6 ،22]. In this study, we employed two distinct types of datasets and conducted 24 experiments to assess the performance of the proposed system comprehensively. The outcomes of these experiments underwent thorough scrutiny. The system architecture is illustrated in Figure 1, encompassing six key steps: (1) Loading the datasets, (2) Data exploration, (3) Dataset preparation, (4) Data splitting, (5) Classification, and (6) Evaluation.

### 5.1 Loading the datasets

Two datasets are utilized in the experiments as follows:

**a. CIC-DDoS2019**

The dataset titled "The Canadian Institute for Cybersecurity (CIC), Distributed Denial of Service Attacks (DDoS)" includes information pertaining to DDoS [40, 41] It consists of 88 attributes, 83 of which were generated by the CICFlowMeter a network traffic flow generator and analyzer and four by. Produced in 2019, this dataset contains approximately 25 GB of classified information. The dataset contains 50 million+ samples, including 56,863 benign and 50,006,249 DDoS attack instances across seven types (LDAP, MSSQL, SYN). It reflects real-world traffic and underwent preprocessing by removing non-informative features like IPs, Flow ID, Timestamp, and SimilarHTTP [41]. The timestamp field was also omitted to encourage learners to focus on discerning stealth and covert attacks rather than relying on time-based predictions. This facilitates easy sample combination or splitting to align with various experimental frameworks. The dataset comprises 45 features of type float64, 37 features of type int64, and 6 features of type object, with a total memory usage of approximately 3.4+ MB. It has a shape of (5000, 82) and a size of 410,000.

**b. CSE-CIC-IDS2018**

This dataset is the outcome of a collaboration between the Communications Security Establishment (CSE) and the Canadian Institute for Cybersecurity (CIC) [42]. Its primary objective is to cater to Intrusion Detection Systems (IDSs) research and incorporates data relevant to cybersecurity incidents. Produced in 2018 [43, 42], it comprises big data with approximately 16,000,000 instances, derived from raw log files totaling about 450 GB [40, 42]. The UNB IDS dataset contains 16.2 million records with 80 features (79 numerical, 1 timestamp) and includes seven attack types: Botnet, Heartbleed, Web Attacks, Brute Force, DoS, DDoS, and Insider Infiltration [43]. The dataset, with dimensions (7000, 80) comprising 560,000 entries, illustrates an imbalanced flow of traffic from 50 attacking machines aimed at 30 servers across 5 departments. Important features consist of Dst Port, Protocol, Stream Duration, and Labels. Its configuration facilitates pattern detection and additional experimentation.

Both the CIC-DDoS2019 datasets and CSE-CIC-IDS2018 demonstrate a notable class imbalance, as there are far more attack instances compared to benign samples (for example, 50 million DDoS attacks against 56,863 benign samples in CIC-DDoS2019). To address this issue, a downsampling technique was implemented to decrease the size of the dominant class, bringing it in line with the minority class for balanced training. This approach facilitated better generalization and lessened model bias. The performance was assessed using F1 Score, Precision, and Recall, showing improved detection of minority instances while having a minimal effect on the accuracy of the majority class.

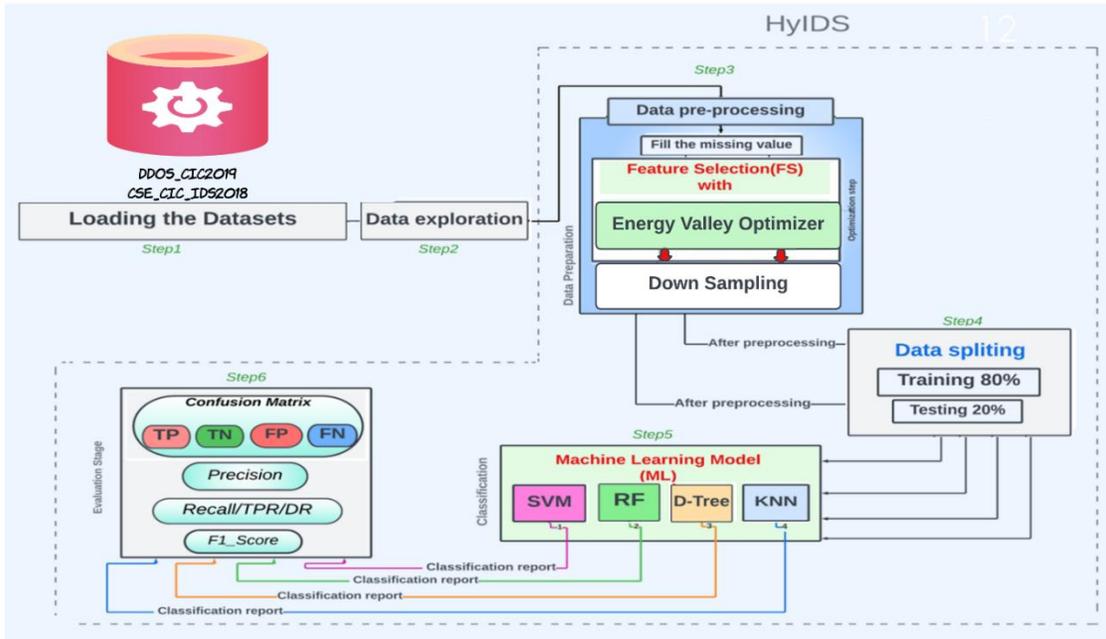

**Figure 1.** The proposed HyIDS Architecture

## 5.2 Data exploration

The exploratory analysis of the CIC-DDoS2019 and CSE-CIC-IDS2018 data constitutes an initial and pivotal phase in comprehending their data structure, dimensions, and features. This process entails scrutinizing various elements such as attack patterns, outliers, features, labels, and values. It is noteworthy that these datasets are characterized by richness, imbalance, large size, realness, tabular format, labeled (supervised), and categorical attributes. The analysis delves into crucial aspects, including attack types, timestamps, source IP addresses, and destination IP addresses. By conducting data exploration, researchers gain profound insights into the datasets, laying a robust foundation for subsequent experiments.

## 5.3 Dataset Preparation

This step encompasses the following:

### a. Data pre-processing

The initial phase involves online data pre-processing of the CIC-DDoS2019 datasets and CSE-CIC-IDS2018. This includes identifying and handling empty or duplicate values across all columns to ensure data quality and reliability. Additionally, the data is uniformly scaled using a min-max scalar to facilitate interpretation and expedite model training. Nominal features are converted into numerical values for ease of interpretation and accelerated model training. Additionally, during the data cleaning process, unrelated fields are removed, and missing values are filled in through machine learning algorithms that estimate them using other attributes within the dataset. These algorithms detect duplicated, noisy, and trivial features that hinder the effectiveness and training of the model. When integrated with machine learning models like SVM, RF, Decision Trees, and KNN, these feature selection methods play a vital role in creating a strong Hybrid Intrusion Detection System (HyIDS) [22, 6, 5]. The adoption of feature selection offers several advantages, including enhanced training speed, reduced complexity for classifiers, improved accuracy, and mitigation of overfitting. Table 4 presents the resultant feature sets of the datasets following the application of EVO for FS, indicating the number of features before and after optimization.

### b. Energy Valley for Feature Selection Method

The EVO algorithm is presented in pseudocode in Algorithm 1. The following steps offer an in-depth description of our innovative approach to feature selection.

This research utilizes the Energy Valley Optimizer (EVO), a metaheuristic approach for selecting features, with the goal of improving the effectiveness of machine learning in detecting intrusions. EVO is inspired by the principle of energy dissipation in potential valleys, working to identify the best feature subsets within complex, high-dimensional datasets. The algorithm starts by generating a set of candidate solutions (feature subsets). It evaluates the performance of each candidate according to their classification outcomes. Through an iterative process of optimization, each candidate adjusts its position using energy-based calculations and interactions with neighboring solutions. The modified candidates' fitness is reassessed, and the one with the highest performance is retained. This process repeats until a predetermined number of iterations is reached or a convergence criterion is satisfied. In the end, the algorithm presents the most appropriate subset of features, striking a balance between reduced dimensionality and improved classification accuracy. Consequently, EVO serves as a potent and scalable technique for discovering the most relevant features in cybersecurity as well as in various real-world datasets. Kindly refer to Table 4 and Figure 2

**Table 4.** Resulted optimal feature set by applying EVO for FS

| Datasets | Features before EVO | Features after EVO |
|---|---|---|
| CIC-DDoS2019 | 88 | 38 |
| CSE-CIC-IDS2018 | 80 | 43 |

**Algorithm 1**: Pseudocode of Energy Valley Optimizer (EVO) for Feature Selection

**Input:** Train data: train = (Xi, Yi) // Training data
  MaxIter = Maximum number of iterations
  PopSize = Population size
**Output:** Best_features    // Optimal selected features
  //**Initialize population and evaluate initial fitness**
1: population = initialize_population (PopSize)
2: fitness = evaluate_fitness (Population, Train_data)
  //**Main optimization loop**
3: **for** iter = 1 to MaxIter **do**
4:   **for** each candidate in a population, **do**
5:     energy valley = calculate_energy (candidate)
6:     neighbors = find neighbors (candidate, energy_valley)
7     update positions (candidate, neighbors)
8:     candidate_fitness = evaluate_fitness (candidate, train_data)
9**:**   **end for**
10:   Best_candidate = select_Best (population, fitness)
11:   **if** stopping_criteria_met then
12**:**     **break**
13**:**   **end if**
14**: end fo**r
  //**Return optimal feature subset**
15: **return** Best_features = Best_candidate. features
16: **end** procedure

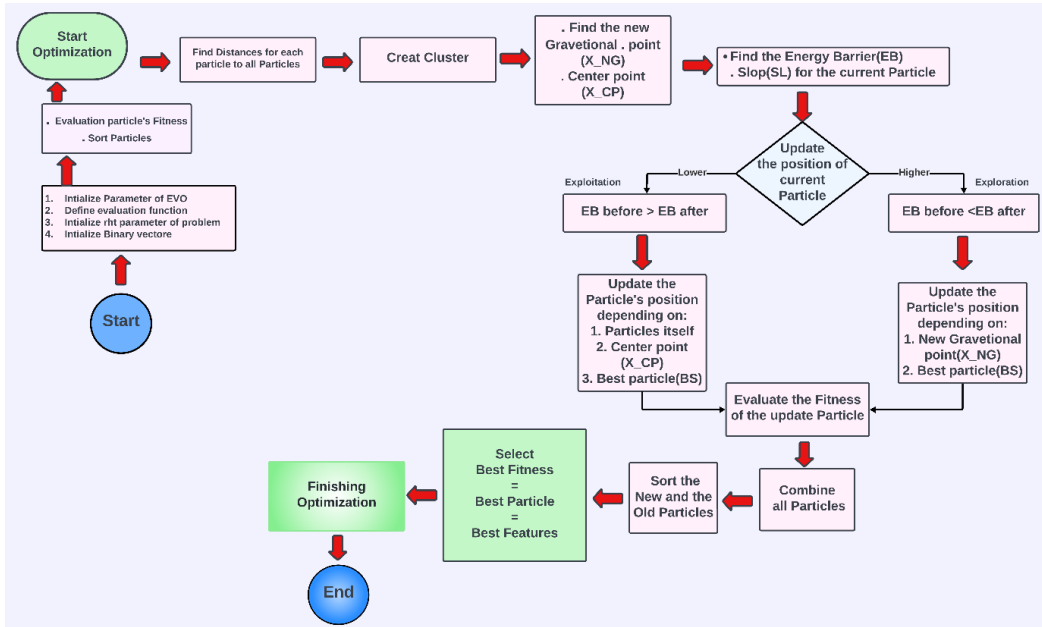

**Figure 2.** The framework of Energy Valley Optimizer for Feature Selections

**c. Downsampling Technique**

The last phase of data preprocessing entails downsampling the imbalanced CIC-DDoS2019 and CSE-CIC-IDS2018 data to achieve equal representation for all classes. Downsampling decreases the number of training samples in the overrepresented class, leading to a balanced dataset and better model performance. This process also involves transforming categorical features into numeric values and selecting 1000 samples from each label. The downsampling method tackles dataset imbalance and boosts classification accuracy. Figures 2(a) and 2(b) show CIC-DDoS2019 and CSE-CIC-IDS2018 datasets before and after downsampling, respectively. Algorithm 2 serves as an illustration to demonstrate the downsampling using the CIC-DDoS2019 data.

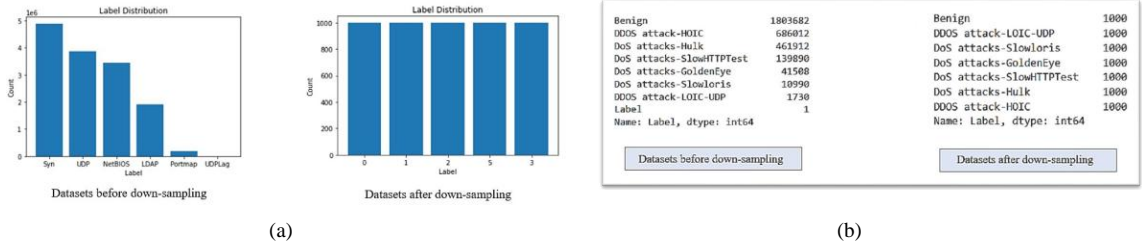

(a)           (b)
**Figure 3.** (a) Before/after downsampling CIC-DDoS2019, (b) Before/after downsampling CSE-CIC-IDS2018

---

**Algorithm 2** Balancing Approach (Downsampling) _CIC_DDoS2019

Procedure Balance for CIC-DDoS2019 Training Dataset
**Input**: An Imbalanced training dataset from CIC-DDoS2019
**Output:** A Balanced training dataset

1. **Initialize** an empty dataset: balanced_training_dataset
2. **Identify** class labels: "Benign", "DDoS Attack"
3. Get instance count for each class:
   • Benign_count = 56,863
   • DDoS_count = 50,006,249
4. **for** each class C in {Benign, DDoS Attack} do
5.     **if** class C is "Benign" then
6.        add all 56,863 instances of "Benign" to balanced_training_dataset
7.     **else if** class C is "DDoS Attack" then
8.        **downsample** "DDoS Attack" to match Benign_count (56,863 instances)
9.        **add** downsampled "DDoS Attack" data to balanced_training_dataset
10.    **end if**
11. **end for**
12. Shuffle balanced training dataset to randomize the instance order
13. **return** balanced training dataset
14. **end** procedure

---

### 5.4 Data splitting

Upon finishing the preprocessing stages, the datasets are divided into two segments for the machine learning algorithms. The initial 80% of the data is allocated for training the model, while the final 20% is set aside for testing. The training dataset is employed to develop the machine learning model, and the testing dataset is used to evaluate the model's effectiveness on unfamiliar data [44, 42, 6].

### 5.5 Classification

We performed ML classification on the CIC-DDoS2019 and CSE-CIC-IDS2018 datasets using SVM, KNN, RF, and D_Tree algorithms to classify data as benign or dangerous. Hybridizing these *ML models* with the *EVO algorithm for feature selection* improves he model performance, identifies important features, improves the accuracy of cloud computing cybersecurity, and obtains the highest accuracy and lowest possible time. After the loop is finished, the global best position obtained from the optimization loop is used to train the final ML model. For the sake of brevity and conciseness, here we present Pseudocode just for the combination of the D_Tree model with EVO for feature selection in Algorithm 3. We utilized four distinct machine learning (ML) models in our research study: (kindly refer to Appendix A)

Numerous supervised machine learning algorithms are commonly used for classification tasks in Intrusion Detection Systems (IDS). The Support Vector Machine with Radial Basis Function (SVM-RBF) is recognized for its capacity to manage nonlinear data by utilizing kernel functions to establish optimal decision boundaries within high-dimensional spaces [45, 46]. Decision Trees (D_Tree) provide an intuitive tree-structured approach to classification by repetitively partitioning the dataset according to feature values, accommodating both categorical and numerical types of data [47, 6]. Random Forest (RF) is an ensemble method that utilizes several decision trees, enhancing both prediction accuracy and robustness via a majority voting process [6, 24]. K-Nearest Neighbor (KNN) is a straightforward yet powerful algorithm that classifies new data points by comparing them to the nearest neighbors in the training dataset [48]. Each of these models has shown impressive performance in IDS applications because of their varied strengths in managing intricate, high-dimensional, or smaller-scale datasets.

## 5.6 Evaluation

For the next step of our analysis and evaluation, the work confusion matrix is used for evaluating the performance of the models [6]. Figure 3 illustrates the confusion matrix along with the explanation of each metric value. Accuracy, Precision, Recall, and F1-score serve as metrics to assess the impact of the ML models on the categorization of attacks. The confusion matrix assesses the classifier's efficacy across three categories. True Positives (TP) and True Negatives (TN) denote accurate predictions, whereas False Positives (FP) and False Negatives (FN) signify erroneous classifications [27, 6]. The metrics used for the evaluation of ML models' performance on the datasets are described as follows:

$$accuracy = (TP + TN) / (TP + TN + FP + FN) \qquad (1)$$
$$precision = TP / (TP + FP) \qquad (2)$$
$$recall = TP / (TP + FN) \qquad (3)$$
$$F1 - score = 2 * (precision * recall) / (precision + recall) \qquad (4)$$

---

**Algorithm 3:** Pseudocode of Decision Tree model Optimized by Energy Valley

---

The effect of utilizing the feature selection method alongside optimization algorithms (EVO) from the next steps:

1: Initialize the parameters:
  - Set and initialize general parameters for EVO, such as the Max. number of function evaluations (MaxFes) and the number of particles (nParticles).
  - Initialize Decision Tree parameters, including Max. depth, Min. samples split, minimum samples leaf, and Max. features.

2: Define the evaluation function evaluate (X, y, features).

3: Initialize problem parameters, such as the number of features (VarNumber), and the lower and upper bounds for binary representation (VarMin and VarMax).

4: Initialize the particles with random binary vectors, where "1" represents the presence of a feature and "0" indicates its absence.

5: Evaluate the fitness of each particle using the evaluation function and store the fitness values in NELs.

6: Sort the particles based on their fitness values. Store the Best Particle (BS) and its fitness value (BS NEL).

7: Define the function distance (a, b) to calculate the distance between two binary vectors.

**Optimization steps**

8: Start the optimization process using the EVO optimizer, which will run until reaching the maximum number of function evaluations (MaxFes):
  (i) For each particle, calculate its distance to all other particles.
  (ii) Find the cluster points (CnPtA and CnPtB) for the current particle based on these distances.
  (iii) Calculate the new gravitational point (X NG) and the center point (X CP) of all particles.
  (iv) Calculate the energy barrier (EB) and slope (SL) for the current particle.
  (v) Update the particle's position based on its energy level compared to the energy barrier (EB):
      (a) If the particle's energy level is lower than the EB, update its position using the particle itself, the center point (X CP), and the Best Particle (BS).
      (b) If the particle's energy level is higher than the EB, update its position using the new gravitational point (X NG) and (BS).
  (vi) Evaluate the fitness of the updated particle and add it to the list of new particles.
  (vii) Combine the new and old particles and their fitness values.
  (viii) Sort the combined particles and fitness values.
  (ix) Keep the top particles to maintain the population size.
  (x) Update the Best Particle (BS) and its fitness value (BS NEL) if a better one is found.
  (xi) Record the Best Fitness at each iteration (BF).

**Classification step**

9: After completing the main loop of the EVO optimization, the Best Particle (BS) represents the best feature subset.

10: Use (BS) to select the corresponding features from the original datasets:
  (i) Use 80% of the data to train the Decision Tree model with the selected features.
  (ii) Use 20% of the data to test the Decision Tree model with the selected features.

11: Evaluate the performance of the model using suitable metrics such as accuracy, precision, recall, and F1-score.

---

## 6 Results and Discussion

In this section, we examine machine learning models used for categorizing data and investigate how metaheuristic algorithms can enhance feature selection. We employ the CSE-CIC-IDS2018 and CIC-DDoS2019 online datasets for this study, using accuracy, recall, and F1-score as metrics for evaluating the performance of the models. This paper aims to improve the security of cloud computing by analyzing the categorization effectiveness of these models.

### a) CIC-DDoS2019 Dataset

This section provides a comprehensive summary of the outcomes obtained from analyzing the CIC-DDoS2019 online datasets using various models. The accuracy results for each model are presented in Table 5. At the same time, Figures 4 and 5 depict the models' accuracies and their durations of both the training and testing phases before applying any FS optimizer, respectively. It is worth noting that, before optimization, the RF model exhibits the highest level of performance among the models. Following RF, the D_Tree model exhibits the next highest performance, with SVM and KNN models performing accordingly. The accuracy results for each model before optimization can be found in Table 5. Table 6 presents the classification results of machine learning models utilizing the optimal feature set derived through the EVO optimizer. For a detailed analysis of the performance enhancements achieved with the EVO optimizer, please refer to Figures 6, 7, and 8. These figures provide in-depth insights into the performance improvements. Moreover, we evaluated the outcomes using the Evaluation Matrix. D TreeEVO emerged as the top performer, boasting a precision of 98.95% and a recall rate of 98.941%. The F1-score, representing the harmonic mean of precision and recall, reached 98.945%. Table 7 and Figure 8 illustrate that the D TreeEVO model significantly outperformed others in terms of F1 score. Conversely, the KNN model lagged with an F1 score of 94.88%. We chose the D TreeEVO model for evaluation due to its unmatched accuracy and performance. This preference is evident in the confusion matrix (Figure 9), where the second, third, and fourth categories demonstrate a preference over others, indicating their offensive nature–NetBIOS attacks (H2), SYN attacks (H3), and UDP attacks (H4). Our analysis showed a classification accuracy of 100% with minimal errors, particularly in accurately categorizing cases for the fourth class. However, a mere 0.0034 cases initially classified as Class four were misclassified as Class one. Notably, the confusion matrix lacks the sixth category, representing Portmap attacks, attributed to the inherent imbalance in the CIC-DDoS2019 dataset. This imbalance leads to a disproportionate distribution of examples across categories, potentially affecting their representation in the confusion matrix, as illustrated in Figure 9. Table 7 presents a comparative analysis between the results derived from our proposed methodology and the findings of prior research conducted on the same dataset, namely CIC-DDoS2019. Previous studies have reported accuracy rates exceeding 94%. In contrast, our proposed method achieved a significantly higher accuracy of 99.26%, demonstrating its superiority over previous studies. Notably, the SVM model with EVO attained a classification accuracy of 95.60%, marking a 1.6% improvement over the highest recorded accuracy in prior research utilizing EVO [48]. Furthermore, the KNN model exhibited a noteworthy enhancement in performance when compared to prior research. The utilization of EVO resulted in improvements of 17% compared to the findings of . Regarding the suggested RF model, it showed a higher improvement rate compared to the work [49] with an increase of approximately 8.5% when utilizing the novel enhancer EVO. This study demonstrates that our proposed method exhibits superior performance compared to related works conducted on the CIC-DDoS2019 dataset.

**Table 5** Classification results based on all features/before optimization

| Model | Accuracy | Precision | Recall | F1-score | Training Time (s) | Testing Time (s) |
|---|---|---|---|---|---|---|
| SVM | 94.66% | 95.11% | 94.19% | 94.31% | 1.12845 | 0.00397 |
| RF | 99.40% | 99.41% | 99.40% | 99.40% | 0.35807 | 0.03033 |
| D Tree | 99.00% | 99.09% | 99.07% | 99.07% | 0.02389 | 0.00219 |
| KNN | 94.93% | 94.88% | 94.89% | 94.88% | 0.00387 | 0.22310 |

**Table 6.** Classification results of models optimized by EVO

| Model | Accuracy % | Precision% | Recall% | F1-score% | Training Time (s) | Testing Time (s) |
|---|---|---|---|---|---|---|
| SVMEVO | 95.60% | 95.40 | 94.89 | 94.99 | 1.10221 | 0.00370 |
| RFEVO | 99.13 | 98.95 | 98.941 | 98.945 | 0.02363 | 0.00347 |
| TreeEVO | 94.93 | 94.88 | 94.89 | 94.88 | 0.003699 | 0.22199 |
| KNNEVO | 95.86 | 95.29 | 95.39 | 1.12832 | 0.00374 | 0.22199 |

**Table 7.** Comparison of the results obtained by our proposed method compared to the existing related works considering the CIC DDoS2019 dataset

| Authors | Model | Proposed method | Accuracy | Precision | Recall | F1-score |
|---|---|---|---|---|---|---|
| Z. Wu et al. [49] | SVM | Robust Transformer (RTIDS) | 94.02% | 94.54% | 94.24% | 94.88% |
| H. Xu et al. [29] | KNN | DL/ BiLSTM | 80.73% | 78.25% | 72.36% | 76.32% |
|  | RF | DL/ BiLSTM | 88.36% | 76.82% | 85.47% | 79.36% |
| Bakro et al. [26] | SVM | GA_GOA | 98.70% | - | - | - |
|  | D_Tree | GA_GOA | 98.99% | - | - | - |
| **Proposed method** | **SVM** | **SVMEVO -FS** | **95.60%** | **95.40%** | **94.89%** | **94.99%** |
|  | **KNN** | **KNNEVO- FS** | **94.93%** | **94.88%** | **94.89%** | **94.88%** |
|  | **RF** | **RFEVO - FS** | **95.40%** | **95.86%** | **95.29%** | **95.39%** |
|  | **D_Tree** | **D_TreeEVO** | **99.13%** | **98.95%** | **98.94%** | **98.94** |

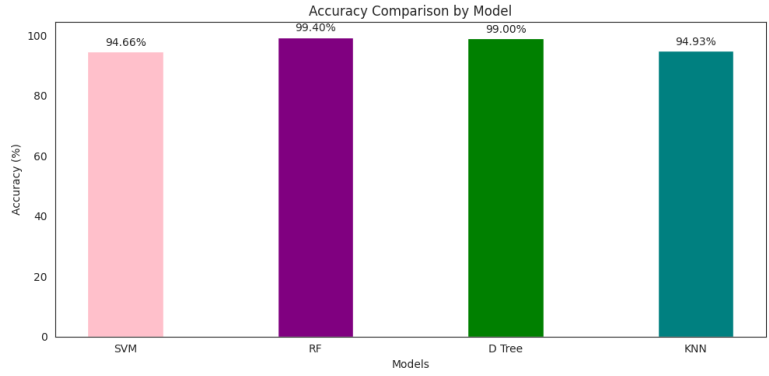

**Figure 4.** Accuracies of models before applying any FS optimizer

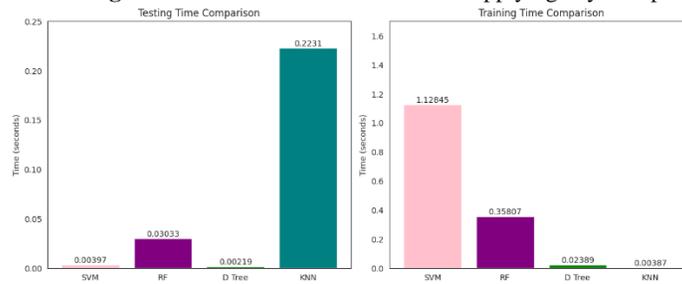

**Figure 5.** Comparison of train/test time lengths of models before applying any FS optimizer

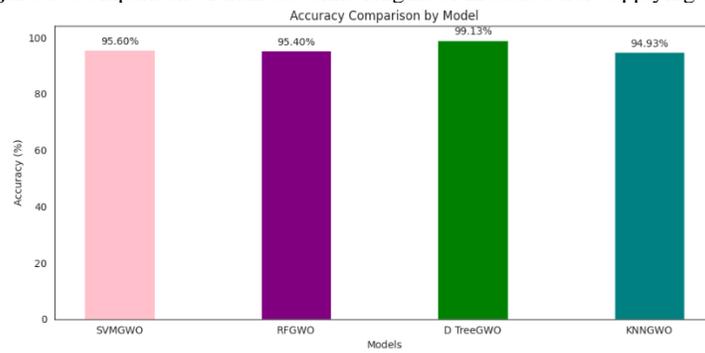

**Figure 6.** Accuracy of models after applying EVO

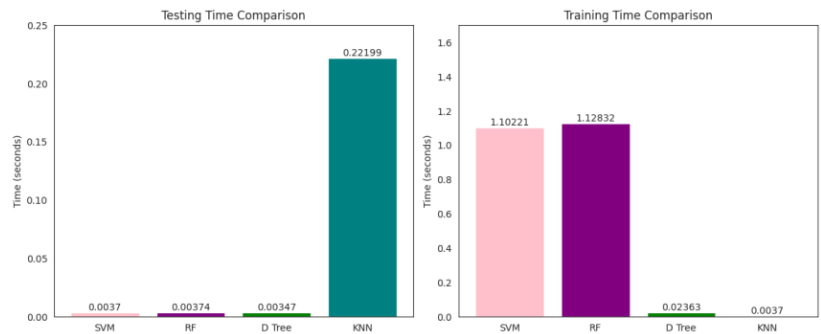

**Figure 7.** Comparison of time for models considering application of EVO

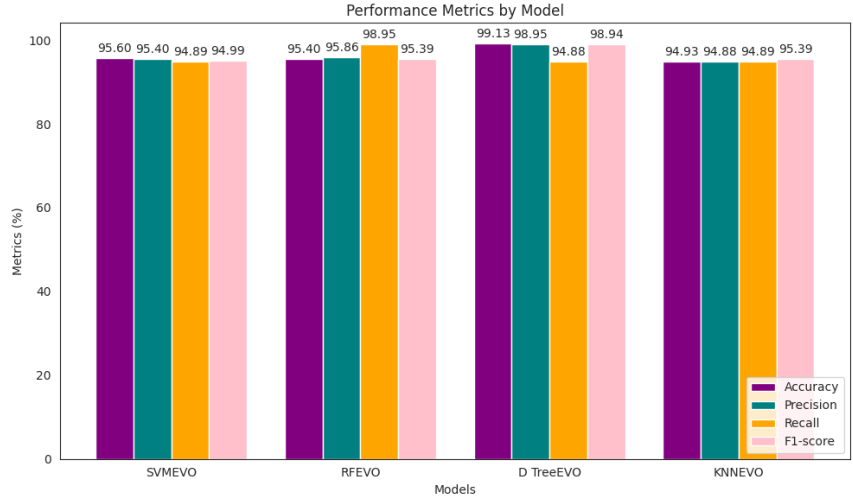

**Figure 8.** Comparison of evaluation metrics for models considering application of EVO

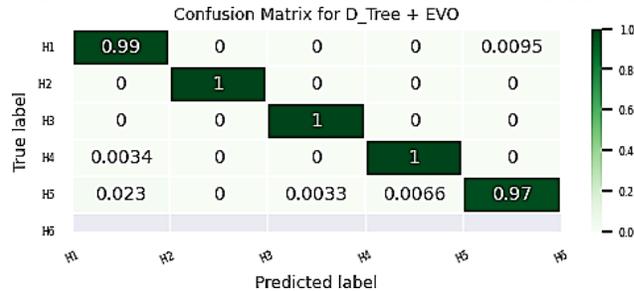

**Figure 9.** Confusion matrix of D TreeEVO

### b) CSE-CIC-IDS2018 Dataset

This section summarizes the results of applying ML models to the CSE-CIC-IDS2018 data, using a selected subset of three files containing various attack types. The selection was made to ensure diverse attack representation while maintaining manageable experimental scope. Tables 8 and Figures 10 and 11 present a detailed overview of the models' performance before optimization. Table 9 and Figure 12 provide a comprehensive summary of the performance enhancements achieved through the utilization of EVO, highlighting the significant improvements obtained through this approach. Figure 13 displays the training and test times of models considering EVO as an FS optimizer. Figure 14 illustrates a comparative examination of evaluation metrics for models employing the EVO algorithm, indicating overall enhancements in accuracy, with the exception of the RF model, which may experience constrained optimization efficacy due to compatibility concerns with the search space and objective function. Significantly, the SVM model attained a 22% improvement, but the Decision Tree model exhibited a 14% augmentation. KNN demonstrated a decrease in training and testing duration by 0.00034 and 0.0135 seconds, respectively. The D TreeEVO method attained the best accuracy of 99.78%, as illustrated in Table 9.

An assessment matrix evaluation confirmed that D TreeEVO achieved the highest precision, true positive rate (TPR), and harmonic mean, all at 99.70%. The confusion matrix displayed in Figure 15 presents the classifications for DDOS attack-LOIC-UDP, DoS attacks-Slowloris, DoS Attacks-GoldenEye, and DoS Attacks-Hulk, with all showing a perfect accuracy of 100%. The D_Tree model produced a remarkable accuracy of over 99.78% and showed a low misclassification rate of 0.00047 for class four. However, it struggled to distinguish between Benign (A1) and DoS-SlowHTTPTest (A5), which adversely affected recall. In Table 10, we present a comparison of our results with previous studies that employed the same models and the CSE-CIC-IDS2018. The study conducted by R. Zhao et al. [10] reported the accuracy rate for all articles. H. Xu et al. [29] reported an accuracy rate of higher than 82%. In contrast, our proposed method achieved an accuracy rate exceeding 97%. This comparison highlights the superior performance of our method about previous studies. The observed increase in performance, about the highest level of accuracy reported in prior research, is 0.2% for EVO as demonstrated in R. Zhao et al. [10]. Our proposed approach reached a precision of 98.56% with EVO, surpassing earlier research. The D_Tree model demonstrated a 12% enhancement compared to the findings of Khan et al. [32] About the suggested KNN model, it demonstrated an enhancement of approximately 17% in comparison to the work conducted by H. Xu et al [29]. With the application of the EVO optimizer, our approach exceeded the performance of prior research conducted on the CSE-CIC-IDS2018. Our findings outperformed earlier studies, including those carried out by H. Xu et al. [29] and Khan et al [32].

**Table 8.** Classification results based on all features/before optimization

| Model | Accuracy % | Precision % | Recall% | F1-score% | Training Time(s) | Testing Time(s) |
|---|---|---|---|---|---|---|
| SVM | 98.28 | 98.57 | 98.50 | 98.51 | 1.62606 | 0.00729 |
| RF | 99.78 | 99.78 | 99.78 | 99.78 | 0.81630 | 0.034129 |
| D Tree | 99.64 | 99.70 | 99.71 | 99.70 | 0.09657 | 0.00339 |
| KNN | 97 | 97.01 | 97.01 | 97.00 | 0.00527 | 0.24064 |

**Table 9.** Classification results of models optimized by EVO

| Model | Accuracy % | Precision% | Recall% | F1-score% | Training Time(s) | Testing time(s) |
|---|---|---|---|---|---|---|
| SVMEVO | 98.50 | 98.56 | 98.50 | 98.51 | 1.60047 | 0.00312 |
| RFEVO | 98.21 | 98.56 | 98.50 | 98.51 | 1.57269 | 0.00429 |
| D TreeEVO | 99.78 | 99.702 | 99.701 | 99.702 | 0.09880 | 0.00338 |
| KNNEVO | 97.00 | 97.01 | 97.01 | 97.00 | 0.00493 | 0.22714 |

**Table 10**. Comparison of the results obtained by proposed method with related works considering CSE-CIC- IDS2018

| Authors | Model | Proposed method | Accuracy | Precision | Recall | F1-score |
|---|---|---|---|---|---|---|
| | RF | DL/ Conv-AE approach | 89.00% | 90.19% | 88.45% | 89.31% |
| Khan et al. [32] | D_Tree | DL/ Conv-AE approach | 89.00% | 77.30% | 82.12% | 79.63% |
| R. Zhao et al. [10] | RF | CFS-DE | 98.01% | 96.61% | 98.01% | 97.30% |
| H. Xu et al. [29] | RF | DL/ BiLSTM | 83.57% | 84.58% | 83.28% | 86.59% |
| H. Xu et al. [29] | KNN | DL/ BiLSTM | 82.86% | 80.53% | 74.22% | 77.29% |
| **Proposed method** | **RF** | **RFEVO - FS** | **98.21%** | **98.56%** | **98.50%** | **98.51%** |
| | **D_Tree** | **D_TreeEVO - FS** | **99.78%** | **99.702%** | **99.701%** | **99.702%** |
| | **KNN** | **KNNEVO- FS** | **97.00%** | **97.01%** | **97.01%** | **97.00%** |

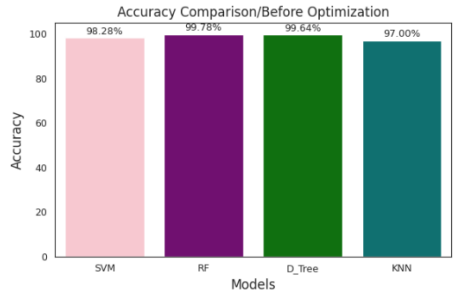

**Figure 10.** Models' accuracies before applying any FS optimizer

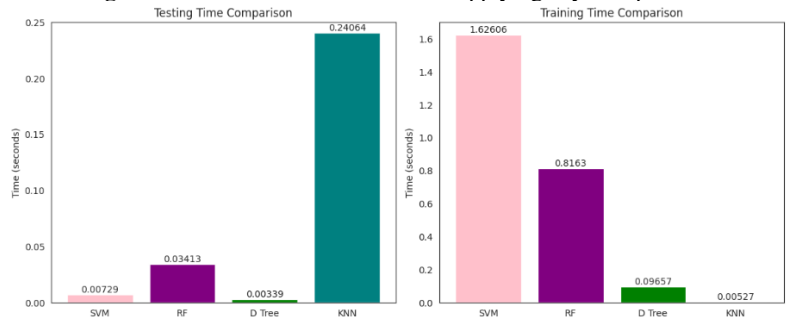

**Figure 11.** Comparison of train/test time lengths of models before applying any FS optimizer

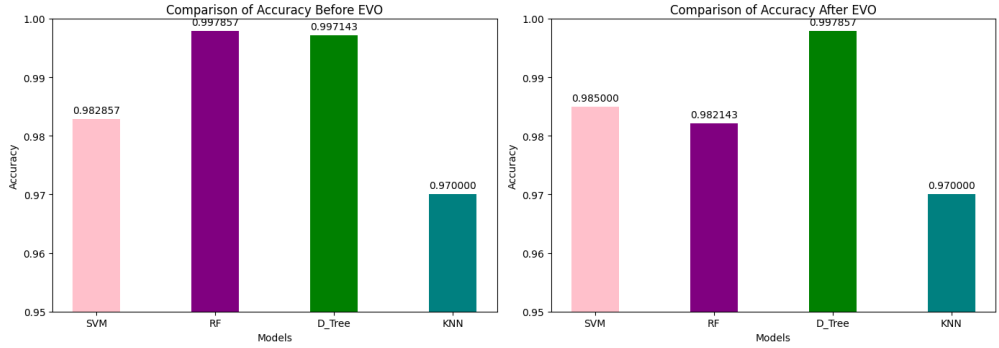

**Figure 12.** Comparison of accuracy before/after applying EVO

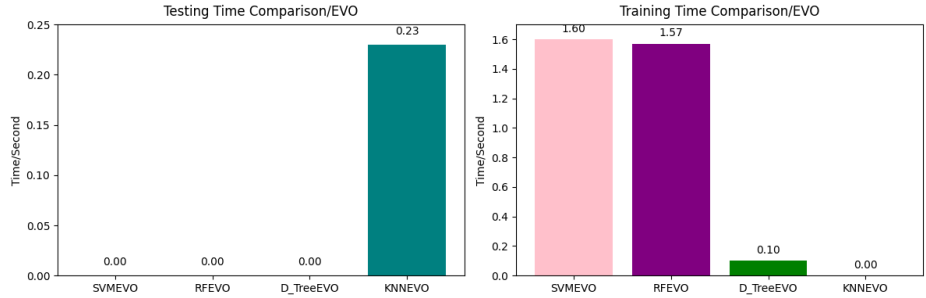

**Figure 13.** Comparison of time for models considering the application of EVO

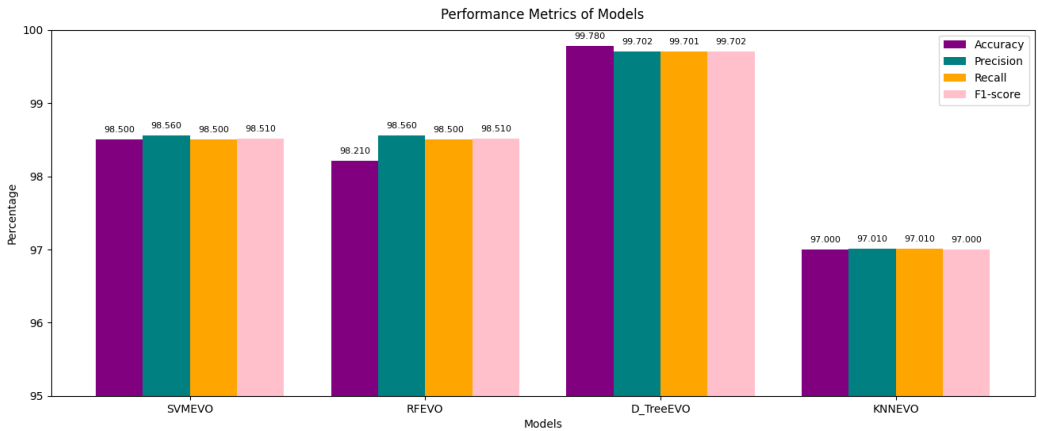

**Figure 14.** Comparison of evaluation metrics for models considering the application of EVO

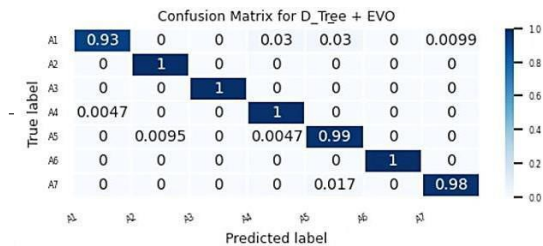

**Figure 15.** Confusion matrix of D TreeEVO

## 7 Conclusions

This study proposes a modern Hybrid Intrusion Detection System (HyIDS) that integrates the novel Energy Valley Optimizer (EVO) for feature selection with machine learning classifiers, including Decision, Support Vector Machine, Random Forest, and K-Nearest Neighbors. The D_TreeEVO model demonstrated high classification performance, achieving 99.13% accuracy on the CIC-DDoS2019 dataset (with 38 features) and 99.78% accuracy on the CSE-CIC-IDS2018 dataset (with 43 features). The results highlight EVO's effectiveness in reducing feature dimensionality while maintaining high classification accuracy, making it a valuable approach for enhancing IDS efficiency. The proposed HyIDS framework offers scalability and adaptability, surpassing traditional IDS techniques. Future research should explore its application to additional datasets, integrate deep learning models, and investigate real-time deployment for enhancing cloud security and intrusion detection mechanisms.

## 8 Future Recommendations

Future studies ought to focus on refining feature selection to address the changing tactics of fraud. The hybrid approach we suggest identifies new features from up-to-date data, thereby increasing the accuracy of HyIDS for detecting large-scale cloud attacks.

## 9 Dataset Sources

a) CIC_DDoS2019 Datasets Available: https://www.unb.ca/cic/datasets/ddos-2019.html
b) CSE_CIC_IDS2018 Datasets Available: https://www.unb.ca/cic/datasets/ids-2018.html

**Appendix A**

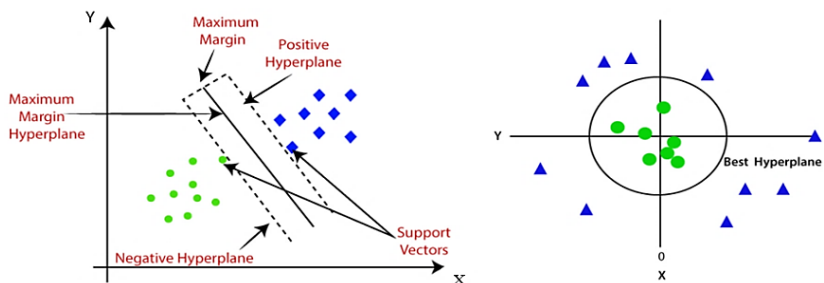

**Figure 16**. Linear SVM and Nonlinear SVM (RBF)[1]

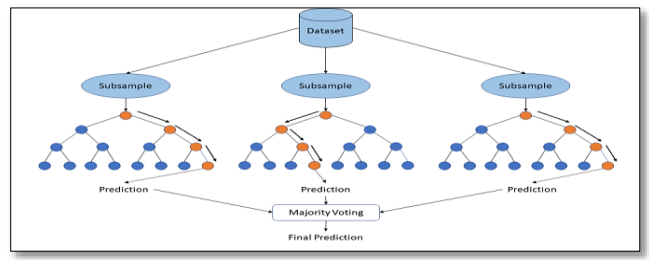

**Figure 17.** An example of an RF algorithm [24]

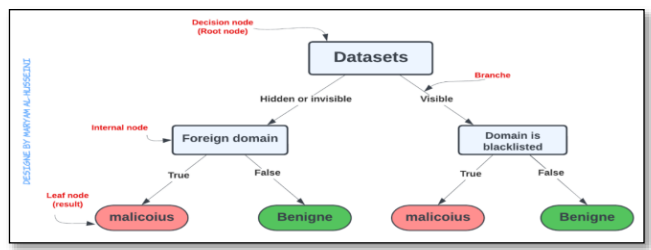

**Figure 18.** An example of D_Tree (Adapted from [22])

---

[1] https://www.javatpoint.com/machine-learning-support-vector-machine-algorithm

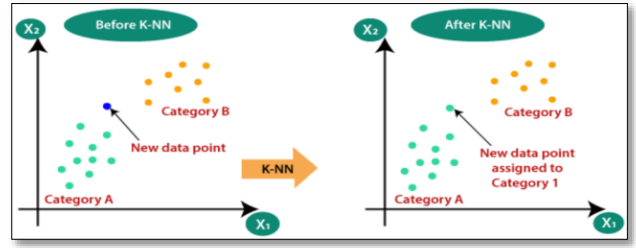

**Figure 19**. Before KNN and After[2]

---

[2] https://www.javatpoint.com/k-nearest-neighbor-algorithm-for-machine-learning